\documentclass{article}
\usepackage{amsmath}
\usepackage{amsfonts}
\usepackage{amssymb}
\usepackage{hyperref}
\author{L.A.\,Melnikovsky\footnote{E-mail: leva@kapitza.ras.ru}}
\title{Quadrupole Stark Effect in Superfluid}
\begin{document}
\maketitle
\begin{abstract}
The roton energy depends on applied electric field gradient due to
its quadrupole moment. This may explain absorption line splitting
recently observed experimentally.
\end{abstract}

Line splitting in the conventional
Stark effect is explained by the interaction of
applied electric field and
intrinsic dipole moment of an atom.
Rotons have no dipole moment
(nor they have any charge).
The leading term in the energy expansion
of a neutral system with zero
dipole moment is given by \cite{LL2}
\begin{equation}
U=\frac{D_{\alpha\beta}}6
\frac{\partial^2 \phi_0}{
            \partial x_\alpha \partial x_\beta}.
\end{equation}
Here $D_{\alpha\beta}$ is the electric
quadrupole moment of the system
and $\phi_0$ is the potential of external
field.

The quadrupole moment of a roton was
estimated earlier
\cite{gravel} as
\begin{equation}
D_{\alpha\beta}
\sim
\frac{Ze\Delta}{24 M^{1/3}
                                c^2 \rho^{2/3}}
\left( \frac{p_\alpha p_\beta}{p^2}-
\frac{\delta_{\alpha\beta}}{3} \right) ,
\end{equation}
where $Z=2$ for helium, $e$ is the elementary
charge, $\mathbf{p}$ and $\Delta$
 are the roton momentum and energy, $M$ is
the atom mass, $c$ is the sound velocity,
and $\rho$ is the fluid density.

Rotons are commonly imagined as microscopic
vortex loops (hence the name). It is interesting
to note that macroscopic
loops are also expected to possess
electric quadrupole moment \cite{nat}.

The frequency shift due to quadrupole
interaction can be estimated as:
\begin{equation}
\Delta f=
\frac{U}{2\pi\hbar}
\sim
\frac{e\Delta}{144\pi\hbar M^{1/3}
                                c^2 \rho^{2/3} l}
E,
\end{equation}
where $l$ is some inhomogeneity length
scale
and $E$ is the electric field. In the
 experiment~\cite{ryba} $l\sim 1\,\text{mm}$.
Thus
\begin{equation}
\frac{\Delta f}{E}
\sim
2\,\frac{\text{Hz}}{\text{V} / \text{cm}}.
\end{equation}
This is close enough
to the experimentally measured value
of $42.65\,\text{Hz}\,\text{cm}/\text{V}$.

To confirm the quadrupole character of
observed effect we propose an experiment
with a different geometry,
so that the influence
of the field \textit{gradient}
on the line splitting can be found.

I thank A.S.Rybalko for fruitful discussion.
I am also grateful to the organizers of
LT25 conference, during which this paper
was written.

\end{document}